# Multiscale Surface-Attached Hydrogel Thin Films with Tailored Architecture

*Benjamin Chollet,[1]\*\* Mengxing Li,[1]\*\* Ekkachai Martwong,[1]*

*Bruno Bresson,[1] Christian Fretigny,[1] Patrick Tabeling,[2] Yvette Tran[1]\**

[1] École Supérieure de Physique et de Chimie Industrielles de la Ville de Paris (ESPCI), ParisTech, PSL Research University, Sciences et Ingénierie de la Matière Molle, CNRS UMR 7615, 10 rue Vauquelin, F-75231 Paris cedex 05, France. Sorbonne-Universités, UPMC Univ Paris 06, SIMM, 10 rue Vauquelin, F-75231 Paris cedex 05, France.

[2] Institut Pierre-Gilles de Gennes (IPGG), 6-12 rue Jean Calvin, 75005 Paris, France.

\* Address correspondence to Yvette.Tran@espci.fr

\*\* B. Chollet and M. Li contributed equally to this work.

**Abstract**

A facile route for the fabrication of surface-attached hydrogel thin films with well-controlled chemistry and tailored architecture on wide range of thickness from nanometers to micrometers is reported. The synthesis, which consists in crosslinking and grafting the preformed and ene-reactive polymer chains through thiol-ene click chemistry, has the main advantage of being well-controlled without addition of initiators. As thiol-ene click reaction can be selectively activated by UV-irradiation (in addition to thermal heating), micro-patterned hydrogel films are easily synthesized. The versatility of our approach is illustrated by the possibility to fabricate various chemical polymer networks, like stimuli-responsive hydrogels, on various plane solid substrates such as silicon wafers, glass and gold surfaces. Another attractive feature is the development of new complex hydrogel films with targeted architecture. The fabrication of various architectures for polymer films is demonstrated: multilayer hydrogel films in which single-networks are stacked one onto the other, interpenetrating networks films with mixture of two networks in the same layer, nanocomposite hydrogel films where nanoparticles are stably trapped inside the mesh of the network. Thanks to its simplicity and its versatility this novel approach to surface-attached hydrogel films should have strong impact in polymer coatings area.

**Introduction**

Hydrophilic polymer coatings that provide specific properties such as wettability, permeability, adhesion or friction properties have drawn considerable attention in various fields, including biomedicine or chemical engineering.[1] The surface properties can be controlled and improved with suitable polymer materials and in particular stimuli-responsive polymers.[2-3] Layer-by-layer polymer assemblies and polymer brushes are so far the most usual polymer coatings. Layer-by-layer assemblies where polymer layers are assembled by physical bonds (hydrogen bond, electrostatic or hydrophobic interactions) have become a universal approach for the facile fabrication of multicomponent films on solid supports. Polymer brushes for which polymer chains are chemically grafted to surface to ensure the durability of the coating are extensively employed in the past two decades. However, layer-by-layer assemblies and brushes as coatings can have some major inconvenience. For polymer brushes, the thickness of polymer brushes which is ruled by the polymer chain length is restricted to low submicrometer. As layer-by-layer assemblies require many cyclic steps, high thickness could be the limitation.

Hydrogel films go beyond and are a real alternative to layer-by-layer assemblies and polymer brushes as stable and durable polymer coatings.[4-5] Hydrogel thin films associate both advantages of hydrogels and films. The architecture of hydrogel films can be inspired from that of macroscopic hydrogels. In recent years, the design of hydrogels has been a hot topic in the research to tailor the structures for well-controlled properties over a large range.[6] For example, the poor mechanical properties of hydrogels materials can be remarkably improved with suitable architectures, such as interpenetrating networks[7] and especially double-networks[8-9] and also hybrid gel materials in which solid nanoparticles are embedded in the polymer network.[10-12] Moreover, the decrease of the feature size in thin films is an appropriate way to create stimuli-responsive hydrogels with fast response. Actually, the response time is roughly expressed as $L^2/D$, where L is the thickness of the hydrogel film and D is the collective diffusion constant of the gel network. It indicates that small size of hydrogels (by downsizing to hydrogel films) would give faster response. For example, the response time is millisecond for micron-thick gel films. The properties of these active materials can be interestingly explored for the fabrication of miniaturized devices with short response time.

Despite their huge potential, there is a lack of clear and universal strategy for the controlled synthesis of surface-attached hydrogel films. We propose a novel, simple and versatile approach to fabricate surface-attached hydrogel films with well-controlled chemistry on wide range of thickness from nanometers to micrometers.[13] This approach must allow easily the adjustment of chemical properties (for example, responsiveness) and physical properties (e.g. size and structure) of the films. Here, we demonstrate the facile route for the fabrication of tailored hydrogel films with responsive properties, the responsiveness providing additional freedom to polymer films. However, the strategy can be extended to more general polymer networks films. Hydrogel films have responsive properties if the polymer chains used for crosslinking are responsive. If mechanical properties are the purpose, it is possible to vary the properties of the polymer network from glassy to rubber.

The strategy chosen for the synthesis of surface-attached chemical hydrogels consists in crosslinking preformed and functionalized polymers by thiol-ene click chemistry. It is preferred to the approach starting from monomers which are then simultaneously polymerized and crosslinked by radical polymerization. In this latter approach, the reaction mixture containing monomers, crosslinkers and initiators is generally confined between two planar substrates (one is functionalized to attach the hydrogel film and the other is non-sticky to detach the film) separated by spacers[14] so that very thin films (submicrometer size) are unfeasible. An additional difficulty is the constraint of controlled atmosphere to avoid oxygen which inhibits radical polymerization. The fabrication of hydrogel films is delicate given that synthesis at surface is very sensitive due to a high surface to volume ratio. To overcome this difficulty, our strategy is to preform polymer chains first and then crosslink and attach them to the surface by click chemistry. Here, thiol-ene click chemistry[15-16] is chosen since the thiol-ene reaction is advantageously performed without any initiator addition and can be activated by either thermal heating or UV-irradiation, resulting in facile patterning of hydrogel films. Many thiol molecules are commercially available to suit the crosslinking reaction and the surface modification: bifunctional thiol (or dithiol) as crosslinkers or gold surfaces modifiers, mercaptosilane to functionalize silicon and glass substrates. The strategy based on thiol-ene click chemistry is likely more straightforward and versatile than the approach using photo-crosslinkable polymers developed by Kuckling et al. with dimethylmaleimide groups [17-21] and Toomey et al. with benzophenone groups.[22-24,25]

In this article, we report the facile route for the fabrication of hydrogel films with responsive properties and tailored architectures. The potential of our approach is demonstrated with different kinds of plane solid substrates, silicon wafers, glass substrates and gold surfaces, and various stimuli-responsive polymers, for example poly(*N*-isopropylacrylamide) for its responsiveness to temperature and poly(acrylic acid) for its pH-sensitivity. We show the synthesis of surface-attached hydrogel films on wide range of thickness from nanometers to micrometers. Spin-coating is chosen as coating technique, but other coating methods can be envisaged, for example dip-coating if thicker films are desired. As discussed, the strategy also allows the creation of patterns if thiol-ene reaction is UV-activated instead of thermally annealed. Another attractive feature of the approach is the development of new complex hydrogel films with various targeted architectures. We show the ability to build multilayer hydrogel films inspired from layer-by-layer assemblies, interpenetrating networks films and nanocomposite hydrogel films which are inspired from the architecture of macroscopic hydrogels. The purpose of this article is to show the facile and universal route for the fabrication of surface-attached hydrogels on external and flat surfaces. The wide possibilities of architecture of the hydrogels are demonstrated: micrometer-size patterns, multilayers, interpenetrating networks, nanocomposite hydrogels. This article is the starting point of our group for many studies on both fundamental aspects and promising applications of surface-attached hydrogels using this synthesis strategy.

**Experimental Section**

*Thiol-modification of substrates.* Thiol-modification of silicon wafers using 3-mercaptopropyltrimethoxysilane is described in detail in a previous article.[13] Briefly, silicon wafers are cleaned in a freshly prepared "piranha" solution before being immersed in a solution of anhydrous toluene with 3 vol% of mercaptopropyltrimethoxysilane under nitrogen for 3 hours. Gold surfaces are obtained by evaporation of a gold film (with thickness around 100 nm) on microscope glass slides. A chromium adhesion layer ($\sim$ 3 nm) is first deposited to improve adhesion of the gold coating. Gold surfaces are functionalized with thiol self-assembled monolayer using a solution of dithioerythritol at 1 mM in chloroform for 2 hours. The thiol-modified surfaces are rinsed in solvent (toluene or chloroform) in ultrasonic bath to remove unreacted thiol.

*Synthesis of hydrogel films.* Ene-functionalized polymers are coated on thiol-modified substrates by spin-coating. A solution containing ene-functionalized copolymers (at various concentration and molecular weight) and dithioerythritol crosslinkers is dropped onto thiol-modified solid substrates (the ratio of dithioerythritol to ene-functionalized copolymer units is 15 times, corresponding to a molar excess of bifunctional dithioerythritol of 30). The conditions of spin-coating are fixed with the final angular velocity of 3000 rpm and the spinning time of 30 seconds. The polymer films are annealed at 120°C for 16 hours under vacuum to activate thiol-ene reaction. For deep UV-irradiation ($\lambda$ = 254 nm) instead of thermal activation, either a common 8 Watt fluorescent lamp with photomasks or a 240 Watt DILASE laser is used. The bilayers films are fabricated using thermal process. The bilayers architecture is made of two stacked polymer network layers which are successively obtained after 16 hours annealing at 120°C. The dry thickness of each layer is equal to 300 nm (obtained using a 254 kg/mol PNIPAM at concentration of 4 wt% in the solution for spin-coating). The total thickness of the bilayers is 600 nm. For the synthesis of interpenetrating networks films, the first network (with dry thickness equal to 300 nm) is synthesized in the same way as single-network films. The formation of the second network (300 nm) inside the first network is obtained by favoring the interdiffusion of chains under saturated atmosphere before crosslinking them. The sample is left under methanol/butanol saturated atmosphere at ambient temperature during 48 hours. It is then put at 120°C during 16 hours to achieve the crosslinking reaction for the formation of the second network. All hydrogel films are rinsed in methanol and sonicated before being dried under nitrogen to remove unreacted polymers so that remained polymers are chemically cross-linked and covalently grafted to the substrate.

Silica-PNIPAM hybrid hydrogel films are prepared by crosslinking polymer chains in the presence of silica nanoparticles dispersed inside. Silica Ludox suspensions (Dupont) which are known for their weak polydispersity were used for the fabrication of hybrid gel films. Ludox-TM50 silica particles diameter of 30 nm (characterized by scanning electron microscopy, dynamic light scattering and small angle neutron scattering) was chosen. Ludox-TM50 is initially purified by dialysis against water during two weeks to remove $Na^+$ stabilizing counter ions (this step is necessary for the addition of the polymer organic solution). After dialysis, a stable suspension is obtained, with the silica concentration around 20 wt%. Ene-functionalized PNIPAM is first dissolved in the mixture of methanol and pure water at a certain concentration, the solution stirred for one night. Then the dialyzed Ludox-TM50 suspension is introduced into the solution whose amount is decided by the ratio between polymer chains and silica particles(with respect to the ratio between methanol and water as V/V = 4/1). Dithioerythritol crosslinkers (with molar excess of 30) are added just before spin-coating. After spin-coating, the samples are immediately put at 120°C under vacuum for 16 hours. The samples are then rinsed in methanol with ultrasonic bath to remove the unreacted polymers and dried with $N_2$ flow.

*Patterns of surface-attached hydrogels.* Patterns are obtained by performing the UV-light irradiation with common 8 Watt fluorescent lamp through a photomask. The photomask is placed over the spin-coated polymer film and the sample is irradiated for 2 hours. Photomasks are fabricated by patterning a 100 nm-thick layer of chromium on quartz plates. Quartz is transparent to UV radiation unlike the 100 nm-thick layer of chromium. To etch the mask design on the chromium layer, a photoresist masking material (Microposit S1813 G2 Photoresist from DOW Chemical) is first deposited according to the design using photolithography. The chromium layer is then chemically wet-etched and the masking material was finally removed. Another way to obtain patterns is to achieve the UV-activation of the thiol-ene reaction with a laser spot which is focused on the surface of the polymer film. The laser spot is shifted over the surface to draw the patterns with very high resolution. This laser lithography is performed using a KLOE DILASE 650 device equipped with a 266 nm laser light source.

*Ellipsometry.* The thickness of hydrogel films is measured with spectroscopic ellipsometer Nanofilm EP3 (Accurion GmbH, Germany). For the measures in air, we use the model with two layers between two semi-infinite media, the silicon substrate (n = 3.87) and the ambient air, the first layer comprising silica and silane (n = 1.46) which thickness was determined before grafting the hydrogel film, the second layer being the hydrogel film which thickness $h_a$ is measured and refractive index is 1.50 (taking in account that hydrogel films contain less than 10% of water in air with humidity ratio between 20% and 60%). *In situ* measurements in water are performed using a temperature-controllable liquid cell with thin glass walls fixed perpendicularly to the light path and the angle of incidence is fixed at 60°. The polymer hydrogel film is modeled as a single layer with the thickness $h_w$ and a constant refractive index between that of water (1.33) and of the polymer (1.52). Since the hydrogel film is covalently attached to the substrate, the polymer amount should keep the same when immersed in water with $n_w = (1.50 - 1.33) \times \phi_p^w + 1.33$ and $h_w \times \phi_p^w = h_a$ with $\phi_p^w$ the volume fraction of polymer in water. The swollen thickness $h_w$, the refractive index of the film in water $n_w$ and the dry thickness $h_a$ are deduced from the fitting of the experimental data. The two equations should provide the same value of $\phi_p^w$ to ensure that the fitting is reliable. Both hydrogel dry films and hydrated films are uniform over a large area (e. g., several centimeters) of the substrates. Each thickness data correspond to the mean measure of at least three samples. For dry films, the standard deviation from the mean value is 7% at the maximum.[13]

The refractive index of the polymer film in water is chosen to be approximated with a linear model instead of Lorentz-Lorenz equation. It is slightly overvalued by the linear model, the divergence being at the maximum 0.1% (for the weakest value of $\phi_p^w$ equal to 0.2), which is well below the accuracy of the refractive index of pure polymer and also much lower than the tolerance accepted for the fitting of ψ and Δ experimental data (1%). Moreover, as the refractive index and the thickness are measured independently by ellipsometry, the value of the thickness is not impacted by this slight divergence of the refractive index.

For hybrid hydrogels, there are three components in the film: polymer, silica and water. So the refractive index in water should be written as $n_w = \sum_i n_i \times \phi_i$ for components *i*. Considering the refractive index of silica equal to 1.46, we have the following system:

$$\begin{cases} n_w = (1.52 - 1.33)\, \phi_p^w + (1.46 - 1.33)\, \phi_s^w + 1.33 \\ h_w \left( \phi_p^w + \phi_s^w \right) = h_a \end{cases}$$

The refractive index of the film in water $n_w$ and the thicknesses of the film in air $h_a$ and in water $h_w$ are deduced from the fitting of the experimental data. The ratio $\frac{\phi_s^w}{\phi_p^w}$ is fixed by the synthesis and corresponds to the silica/polymer ratio shown in the Supporting Information. The equations should provide consistent values of the volume fraction of polymer and silica in the hybrid gel film, $\phi_p^w$ and $\phi_s^w$, so that the fitting is reliable.

*Atomic Force Microscopy.* All AFM images are acquired with a Bruker ICON microscope controlled by a Nanoscope V. Height and phase images are obtained in tapping mode using a tip with a spring constant of 40 N/m and a resonance frequency close to 300 kHz. The working conditions are the following: free amplitude of about 20 nm, amplitude reduction equal to 0.95, scan frequency of 0.5 Hz and images acquired with 512 x 512 pixels. We check that all the images are acquired in repulsive mode. The phase difference between the cantilever excitation at its resonance and the response of this cantilever are measured by the internal ICON lock-in amplifier. Both height images and phase images are displayed after baseline subtraction so that the height and phase values are relative values. The mean roughness is estimated from the half width of the height histogram of each image.

## Results and Discussion

**Strategy for the synthesis of hydrogel thin films**

Surface-attached hydrogel films are synthesized by simultaneously crosslinking and grafting responsive polymers by thiol-ene click reaction. The preformed and ene-reactive polymer is coated on thiol-modified substrate with dithiol crosslinkers (Figure 1). The preformed polymers provide responsive properties of hydrogel films. The synthesis of ene-functionalized polymers can be carried out in two steps for ene-functionalized PNIPAM (one step for PAA) as described in detail in a previous paper.13

First, P(NIPAM-*co*-AA) copolymers are synthesized by free radical polymerization (this step is not necessary for PAA homopolymer which can be easily and cheaply purchased). Since there is no need to control the polydispersity index of polymer chains which are aimed to be post-crosslinked, free radical polymerization is the best option, as it is much straightforward compared to other polymerization methods. The initiator used for radical polymerization in water is ammonium persulfate /sodium metabisulfite redox couple. By varying the concentration of the reducing agent, polymer chains can be obtained with different molecular weight, as shown by Bokias et al.[26] Second, the copolymer chains are randomly functionalized with ene-groups by peptide reaction. Because there are carboxyl groups (-COOH) in the polymer chains and allylamine has both amino (-NH$_2$) and ene-group, the peptide reaction is exploited to graft allylamine onto polymer chains in the presence of 1-(3-dimethyl-aminopropyl)-3-ethylcarbodiimide hydrochloride (EDC) and *N*-hydroxysulfosuccinimide (NHS). EDC is used as the dehydration agent and NHS as the addition agent to increase yields and decrease side reactions.[27] Both radical polymerization and peptide reactions are performed in water for a facile recovery of the ene-functionalized polymer after purification by dialysis against water and freeze-drying. However, if the purification is necessary for the characterization of the ene-functionalized polymer (SEC, NMR), it is not required for the synthesis of chemical hydrogel film grafted to solid substrate. The washing of surface-attached film is enough to remove all undesired species such as residual monomers, initiators, reactive products and byproducts from synthesis of ene-functionalized polymer.

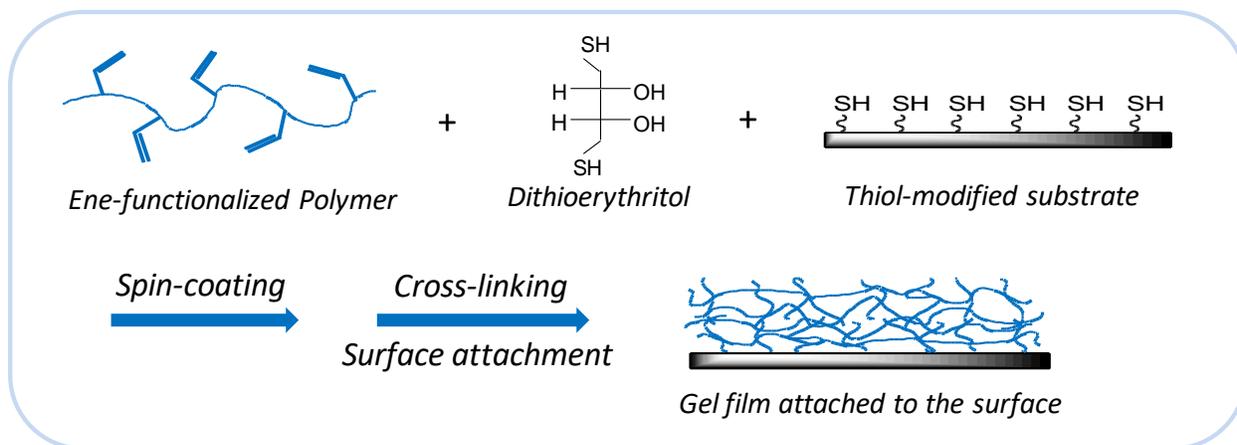

*Figure 1.* Schematic of synthesis of surface-attached hydrogel films: ene-functionalized polymer is spin-coated on thiol-modified substrate with dithiol crosslinkers. Thiol-ene click reaction allows simultaneous crosslinking and surface attachment of polymer chains.

Hydrogel films are grafted on various solid substrates such as silicon wafers, glass substrates and gold surfaces. Thiol-modification of these substrates is rather common by using commercial thiol molecules. Silicon wafers and glass substrates are thiol-silanized with mercaptopropyltrimethoxysilane and gold surfaces with dithiolerythritol which is also used as crosslinkers (other dithiol molecules are also suitable). There are many coating methods to get homogeneous (melt) polymer films from polymer solutions such as dip-coating, spray-coating or spin-coating. We chose the spin-coating technique to synthesize hydrogel films on wide range of thickness from nanometers to micrometers. The spin-coating technique has also the advantage to necessitate a little polymer. For example, as only 0.2 ml of solution is needed to cover a surface of 1 cm$^2$, a polymer concentration of 1 wt% requires only 2 mg of polymer. The solvent blend selected to prepare the polymer solution must solubilize both ene-functionalized polymer and dithiol crosslinkers and also allow the spreading of the polymer film on thiol-modified substrates by spin-coating. The mixture of butanol and methanol (V/V = 1/1) is preferred as the solvent blend for PNIPAM and the mixture of methanol and formic acid (V/V = 7/3) for PAA.

The film thickness is strongly affected by many factors such as the volatility of the solvent, the final angular velocity and the viscosity of the polymer solution.[28] Here, films with different thickness are obtained by changing the viscosity with the variation of polymer molecular weight and polymer concentration. Figure 2 demonstrates that surface-attached hydrogel films can be synthesized on wide range of thickness from nanometers to micrometers. As expected, the film thickness increases with the viscosity of polymer solution, and so with polymer concentration or molecular weight. For molecular weight of 66 kg/mol, submicrometer thicknesses can be obtained with high precision (below 5% deviation) by varying the concentration below 10 wt%. For higher molecular weight, the films have a bigger deviation of the thickness but the advantage is the large range with the formation of micrometric films. The results are shown here for PNIPAM layers with various molecular weight but the same are obtained for PAA films. In particular, we observed that the film thickness is similar for PNIPAM and PAA using spinning polymer solutions with same concentration, same molecular weight and suitable solvent (see Supporting Information).

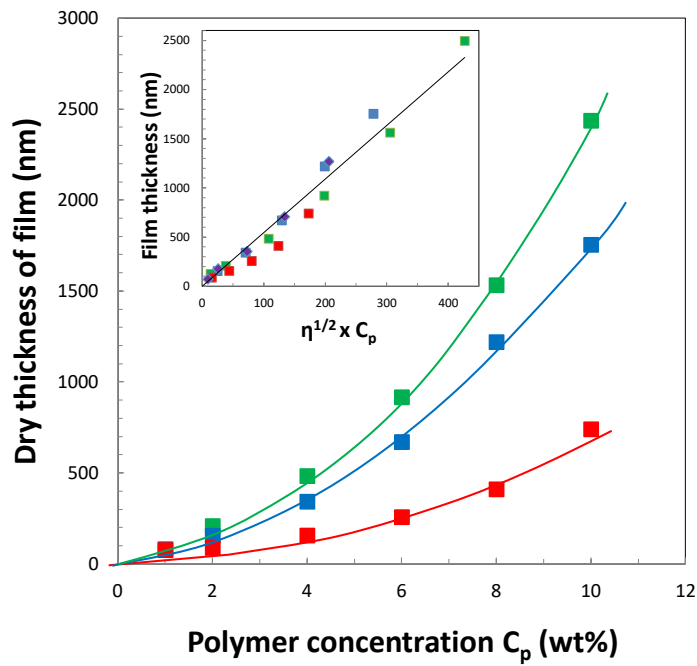

*Figure 2.* Dry thickness of PNIPAM hydrogel films as function of polymer concentration in the spin-coating solution for various molecular weight (66 kg/mol: red markers, 254 kg/mol: blue markers, 669 kg/mol: green markers, the solid lines being guides for the eye). (Insert) Thickness of hydrogel films as function of $\eta^{1/2} C_p$ or $M^{\alpha/2} C_p^{3/2}$ with $\alpha = 0.8$. The data are from PNIPAM hydrogel films with various molecular weight (same color code as the principal figure). Also are shown the data for PAA hydrogel films with 250 kg/mol (purple data), the thickness being the same as that obtained with PNIPAM film with 254 kg/mol.

The dependence of the film thickness on the viscosity of the polymer solution is clearly demonstrated in the insert of Figure 2. In the spin-coating process, the film thinning can be modeled by hydrodynamic equations describing radial out-flow and solvent evaporation.[29] The thickness of ultrathin polymer films can be expressed by the scaling: $h \propto \omega^{-1/2} \eta^{1/2} C_p$ where $h$ is the final film thickness, $\omega$ is the spin speed, $\eta$ is the initial solution viscosity and $C_p$ the polymer concentration. The power law dependence of the viscosity with the exponent of 1/2[30] is preferred to the value of 1/3[31] assuming that all the solvent flashes off at a certain time. This assumption is in agreement with our spin-coating conditions with the use of very volatile solvent such as methanol. As the viscosity of polymer solution can be expressed as function of the concentration and the molecular weight as $C_p M^\alpha$ (Mark-Houwink parameter $\alpha$ is between 0.5 for theta solvents and 0.8 for good solvents). Finally, the thickness of polymer films is given by the expression: $h \propto \omega^{-1/2} M^{\alpha/2} C_p^{3/2}$. The film thickness is plotted as $M^{\alpha/2} C_p^{3/2}$ using $\alpha = 0.8$. Experimental data are satisfactorily aligned on the same master curve on a large range of concentration (from 1 to 10 wt%) and large range of thickness from nanometer to micrometers. Note that in our spin-coating experiments, the spin speed was fixed at 3000 rpm, but the variation of this parameter could be as valuable as that of polymer concentration and molecular weight to reach the desired thickness.

The formation of surface-attached hydrogel films can be achieved by either UV-irradiation or thermal activation. Deep selective UV-irradiation allows the activation of thiol-ene reaction without any initiator to simultaneously crosslink polymer chains and graft them on thiol-modified substrate. Figure 3 shows PNIPAM film thickness as function of the irradiation time. The experiments were performed using a commercial 8 Watt fluorescent lamp at a wavelength of 254 nm. The film thickness increases with the exposure time and reach a plateau after 2 hours. The maximum thickness reached for each polymer concentration (in the solution for spin-coating) is the same than that obtained by heating at 120°C for 16 hours. The variation of the film thickness with the exposure time during the first two hours is probably due to partial crosslinking of polymer chains, the uncrosslinked chains being removed by washing. The same kinetics is observed for PAA surface-attached hydrogel films, as shown in Supporting Information.

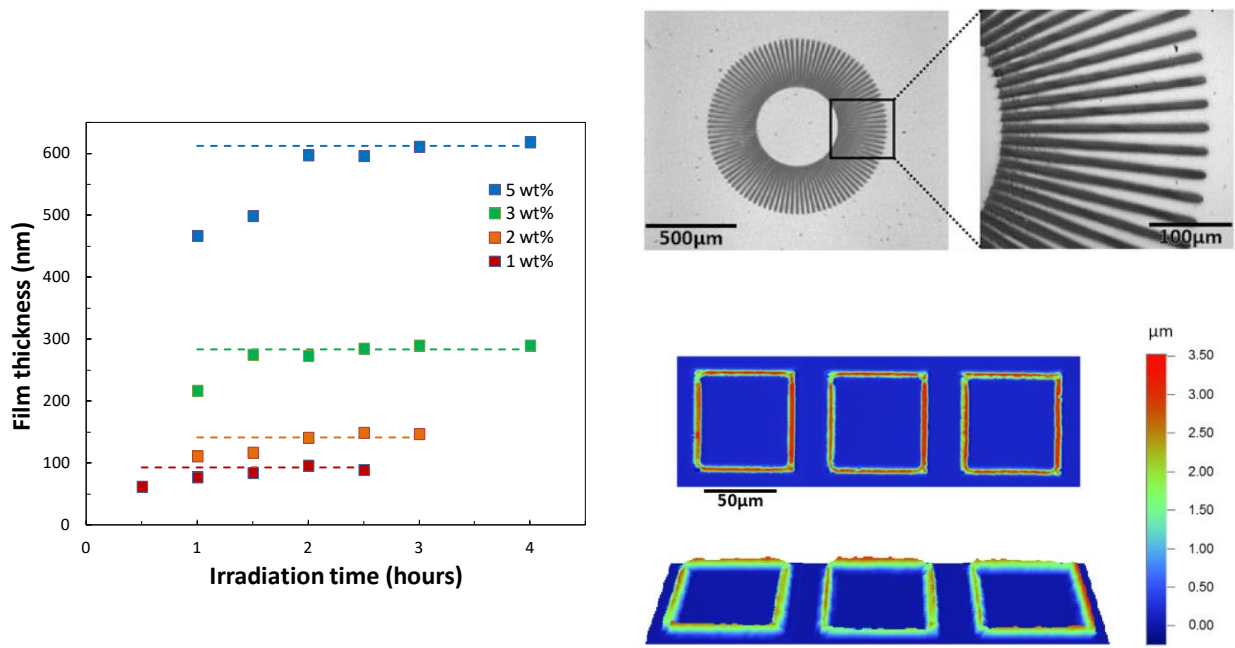

*Figure 3.* (Left) Dry thickness of PNIPAM hydrogel films as function of UV-irradiation time for various concentrations of polymer in the solution for spin-coating (with molecular weight of 254 kg/mol). The dotted lines indicate the maximum thickness of hydrogel films which is also obtained by heating activation. (Top-right) Optical microscopy image of patterned PNIPAM hydrogel (with dry thickness equal to 250 nm) on silicon wafer. The patterns were obtained by using 8 Watt fluorescent lamp with photomask. (Bottom-right) Top view and 3D view of patterned PNIPAM hydrogel (with dry thickness equal to 3 μm) on glass substrate. The patterns were obtained by using 240 Watt laser technology such as DILASE650 device. Height profiles were acquired with a Wyko NT9100 optical profilometer.

In addition to time saving, the advantage of the UV-activation is the ability to create patterns of surface-attached hydrogels. The patterning of hydrogels can be obtained using common 8 Watt fluorescent lamp with photomasks. This patterning process corresponds to a negative photolithography. It can also be achieved without photomask by using 240 Watt laser technology such as DILASE650 device. As a simple illustration, Figure 3 shows a clear patterning of PNIPAM hydrogels on silicon surface with 6 microns-wide lines using the fluorescent lamp through a photomask. Images of the patterns were obtained with a reflected light microscope. The reflective silicon surface appears bright and the hydrogel patterns appear in black tones. Figure 4 also displays squared patterns achieved from direct writing with the DILASE laser device. Height profiles were obtained with an optical profilometer, the images being in false colors. The glass substrate is shown in deep blue, and the gel is colorized from green to red depending on the local thickness. It clearly demonstrates that patterns of surface-attached hydrogels can be drawn on solid substrates with high resolution. This facile micro-patterning of surface-attached and responsive coatings with well-controlled chemistry is very attractive for many applications. Swollen hydrogel patterns (in water) do not display morphological instabilities resulting from the hydrogel buckling when hydrogel patterns synthesized in dry state are exposed to water. These buckling instabilities are due to inhomogeneous compressive strains that arise from the geometric constraints.[25] Actually, our surface-attached hydrogel patterns are rather wide than thick (the thickness is in general ten times lower than the width) so that the swelling of hydrogel is unidirectional without out-of-plane deformation. Hydrogel patterns will be the subject of a forthcoming paper and in particular the fabrication and characterization of patterns with multi-responsive properties will be focused.

**Stimuli-responsive properties of hydrogel films**

Temperature-responsive properties of PNIPAM hydrogel films are demonstrated with the measure of the thickness of hydrogel films in water at different temperature (Figure 4). The swelling ratio is defined as the ratio between the thickness of the hydrogel film in water (swollen thickness) and in air (dry thickness). Indeed, from the measure of water content in PNIPAM hydrogel film with controlled humidity ratio (data not shown here), PNIPAM hydrogel films are shown to contain less than 10% of water in air (for humidity ratio roughly varying from 20% to 60%). For simplicity, the thickness of PNIPAM hydrogel film in air is assumed to be the dry thickness. The swelling ratio is independent of the thickness in the whole range from 100 nm to a few micrometers. The data were obtained with hydrogel films of supposedly same crosslinks density. This assumption seems consistent since the synthesis of the hydrogel films is performed with the same (2%) reactive copolymer and same (30 times the ene-reactive groups) excess of dithioerythritol crosslinkers. As experimental conditions used the same except the thickness, hydrogel films can be supposed to be chemically the same. The crosslinking is likely uniform without any gradient in the direction normal to the surface as ene-functionalized polymers and dithiol crosslinkers are homogeneously mixed before spin-coating (both are totally soluble in the solvent used). Actually, it was shown that surface-attached hydrogel films have abrupt density profile with weak interface width. Experimental data from ellipsometry and neutron reflectivity could be satisfactorily fitted with a model of one unique and uniform layer for PNIPAM hydrogel films.[13] Unfortunately, spectroscopic methods probing surfaces are unable to provide information on the crosslinks density. For example, XPS technique only probes nanometric layers (the standard penetration depth of the technique is about 5 nm) and cannot give average measure of the whole hydrogel films. Infrared spectroscopy in ATR (Attenuated Total Reflection with micrometric penetration depth) is more suitable to probe the whole films. However, the spectra analysis is problematic for wave number range below 1600 cm$^{-1}$ due to the high absorbance of silicon substrates used as (infrared) waveguides. Also, the ratio of thiol-ene reaction cannot be quantified as the absorption peak which is characteristics of S-C bond (around 700 cm$^{-1}$) is in this range of low wave number. Moreover, the thiol-ene ratio is weak with a maximum of 2%, the ratio of ene groups measured by $^1$H NMR being 2% (see in Supporting Information).

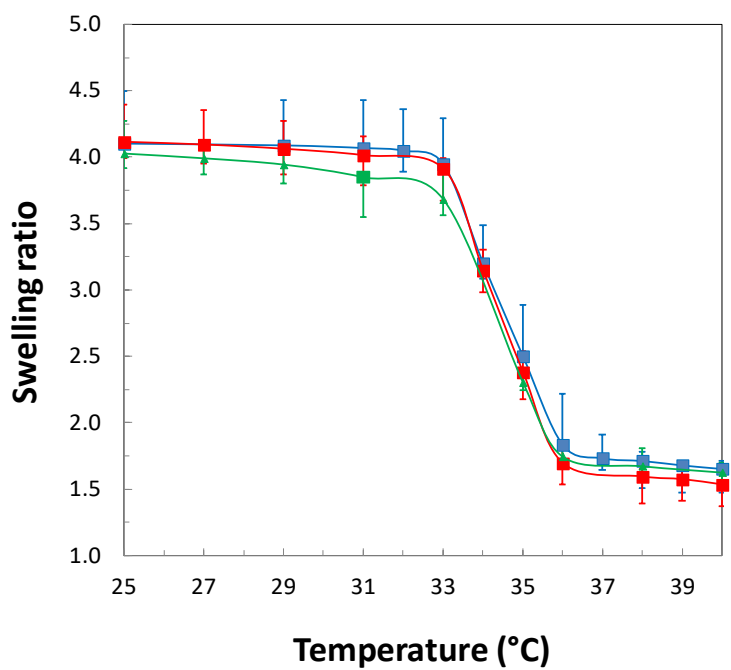

*Figure 4. Swelling ratio of PNIPAM hydrogel films as function of temperature for various dry thickness of the films, using end-functionalized PNIPAM with the molecular weight equal to 254 kg/mol. Blue data: 150 nm, red data: 250 nm, green data: 420 nm.*

The measure of the thickness in the swollen state is a way to control the synthesis of hydrogel films since it provides (indirectly) the crosslinks density. The weaker the crosslinks density, the more the hydrogel films swell. Differently from macroscopic hydrogels, surface-attached polymer networks are expected to swell linearly, i.e. only in the direction normal to the substrate and not in the direction parallel (the dimension of the surface is infinite compared to the perpendicular direction). The Flory-Rehner theory[32] for the swelling of hydrogels extended to one-dimension swelling is appropriate, as shown by Toomey et al.[22] They determined the swelling ratio (or linear degree of swelling) as function of the proportion of photo-crosslinkable units from 0.5% to 14.3% for 100 nm-thick films. Our swelling ratio of 4 found with 2% ene-reactive polymer is in agreement with their results.

If the swelling ratio is 4 for temperature below the LCST, the value of 1.5 is found for temperatures above the LCST. It means that collapsed PNIPAM hydrogel films are not free of water but indeed contains roughly 30% of water. It is in good agreement with the results from Kuckling's group[17-18] and Toomey et al.[23] They also found that water is not entirely expulsed from collapsed PNIPAM hydrogels no matter the crosslinks density. The swelling-collapse phase transition of responsive hydrogel films with swelling ratio between 4 and 1.5 is quite high amplitude (almost 3), which is interesting for applications exploiting temperature-responsive properties. This amplitude of swelling-collapse transition was also found by Toomey et al.[23] The amplitude could be increased with weakly cross-linked hydrogels. However, weakly cross-linked hydrogels (with high water content) are not mechanically stable and can irreversibly be damaged under constraints due to the swelling. Polymer chains are then removed during the washing step. Another point which could be valuable for applications is the sharp transition within a temperature range of 3°C (here from 33°C to 36°C). The identical sharp transition with high amplitude is found for surface-attached PNIPAM films in the whole range of the thickness from 100 nm to 1 μm. In a previous publication, we have also demonstrated that the swelling ratio which only depends on the cross-links density is the same for a few micrometers-thick films.[13]

**Mulilayers hydrogels and interpenetrating networks**

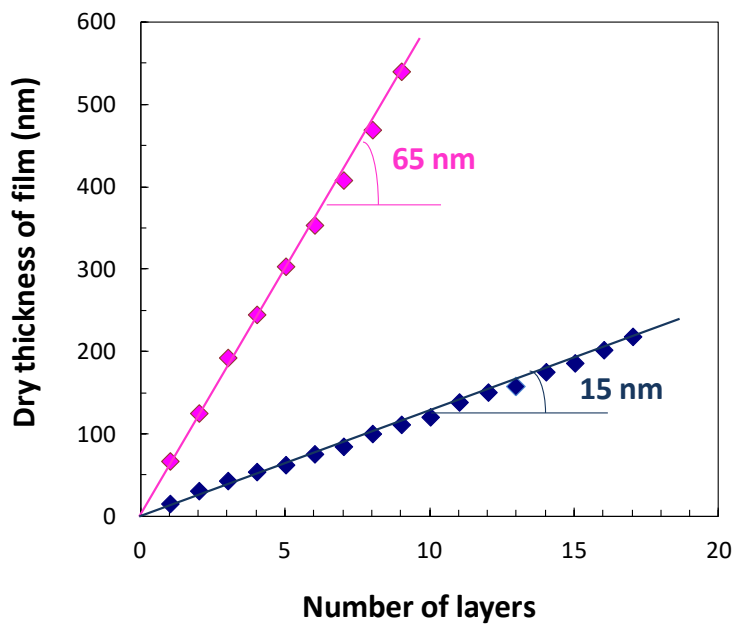

*Figure 5.* Dry thickness of PNIPAM hydrogel multilayers film as function of the number of layers. The linear increase is shown for two thicknesses of the single layer, the slopes being equal to 15 nm/layer (blue data) and 65 nm/layer (magenta data). Multilayers are obtained with successive spin-coating/heating of PNIPAM layers.

Beyond single-network films, we also develop new hydrogel films with various targeted architectures. We fabricate multilayers hydrogel films inspired from layer-by-layer assemblies in which single-network is stacked one onto the other and interpenetrating networks films with the mixture of two networks in the same layer. Multilayers gel films are available by successive deposition of a new layer on top of the former layer, the thickness of each layer being controlled by the molecular weight of the polymer and the concentration of the solution for coating. The very first layer is synthesized like a single-network film. The preparation of the next layers is carried out with the same process: coating of polymers and crosslinking by thiol-ene reaction.

The dry thickness of the hydrogel films (or thickness measured in air) increases linearly with the number of deposited layers (Figure 5). The linear increase is demonstrated for two different thickness of the elementary layer (15 nm and 65 nm) decided by the concentration of the spin-coating solution as shown previously in Figure 2. The hydrogel can be any polymer PNIPAM or PAA. It provides evidence the simplicity of the method to fabricate stable and durable multi-responsive polymer films. The potential of multilayers hydrogel films is the chance to vary the thickness at will in comparison to layer-by-layer polymer assemblies in which the elementary adsorbed layer is a few nanometers-thick due to the obligatory physical interaction between two adjacent layers.

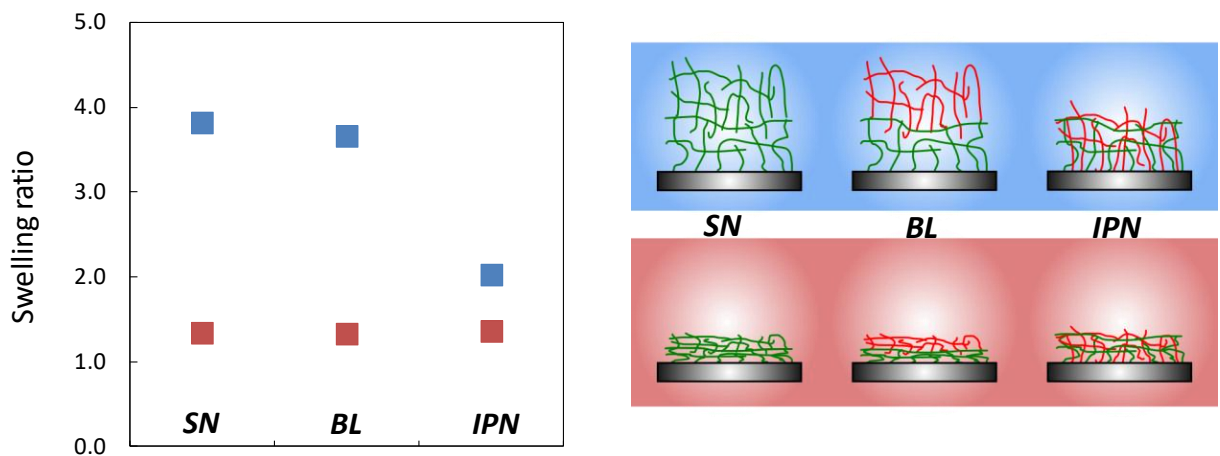

*Figure 6*. (Left) Swelling ratio of PNIPAM hydrogel films for various architectures: SN for single network, BL for bilayers and IPN for interpenetrating networks. The data are shown for 25°C in the swollen state (blue markers) and 40°C in the collapsed state (red markers). (Right) Schematic of hydrogel films swelling for various architectures (SN, BL and IPN) in the swollen state (in water at 25°C, top) and in the collapsed state (in water at 40°C, bottom).

For interpenetrating networks, the mixture of two networks is targeted. The strategy is accordingly different from that used for multilayers films. After the formation of the first network, the polymer chains coated for the second network need to diffuse inside the first network before being crosslinked. It is not possible with the dry thickness of films to differentiate the architecture of bilayers BL and interpenetrating networks IPN as the dry thickness (measured in air) corresponds to the amount of polymer. However, a bilayers film and an IPN film made from the same single-network (with same polymer amount and same crosslinking ratio) are expected to swell differently. The swelling ratio at 25°C of a bilayers film is expected to be the same as the ratio of single-network film and twice the ratio of IPN film because the whole IPN film can be considered as twice crosslinked due to the interdiffusion of the two networks.

Figure 6 displays the swelling ratio of PNIPAM hydrogel films in water at 25°C (below the LCST) and 40°C (above the LCST). The results are shown for various architectures such as single-network, bilayers and interpenetrating networks. All hydrogel films were synthesized using thermal activation for crosslinking and surface-attachment. (i) The single-network films used as the reference have two various dry thicknesses: 300 nm and 600 nm. (ii) The bilayers architecture is made of two stacked layers which are successively coated and crosslinked. The dry thickness of each layer is equal to 300 nm providing a total thickness of 600 nm. (iii) For the fabrication of interpenetrating networks film, the first network (with dry thickness equal to 300 nm) is synthesized in the same way as single-network film. The polymer chains for the formation of the second network are then coated with dithiol crosslinkers on top of the gel film (the dry thickness of the second film is also equal to 300 nm). The total interdiffusion of chains inside the first network is favored by the swelling or stretching of the first network under vapor saturated atmosphere (methanol/butanol for PNIPAM film). Finally, the thermal process allows the crosslinking reaction for the formation of the second network. At 25°C below the LCST of PNIPAM, the swelling ratio of bilayers film is the same as that of single-network film, as expected. The swelling ratio of 4 is also the value found in Figure 4. It clearly proves architecture of stacked layers with very little interdiffusion of chains between the layers. The stability of bilayers is probably ensured by entanglements between peripheral chains and thiol-ene bonds between the chains of the first layer and those of the second layer. The swelling ratio of IPN film is half the swelling ratio of single-network film and bilayers film since the whole

IPN film can be considered as twice crosslinked due to the complete interdiffusion of the two networks. At 40°C above the LCST of PNIPAM, in the collapsed state, the swelling ratio is around 1.5 for all films whatever the architecture. Collapsed PNIPAM gel films are not free of water while containing approximately 30% of water independently of the architecture and the thickness of the gel films, as discussed previously.

Here, we show a way to fabricate interpenetrating networks film by successively coating reactive polymers of the two (or more) networks. It should be noted that the other approach to interpenetrating networks films could be the use of two (or more) distinctive and selective chemical reactions to simultaneously crosslink the two (or more) networks.[33] The advantage is the simultaneous coating of the two polymers for the formation of interpenetrating networks.

**Surface-attached hybrid hydrogel films**

Another example of architecture of hydrogel films with great potential is the hybrid gel films in which nanoparticles are incorporated inside the network. An illustration of nanocomposite hydrogel films is given with silica solid particles stably trapped inside PNIPAM hydrogel. The nanocomposite networks are achieved by simply adding the silica nanoparticles to the solution of polymer for coating. The silica particles used in this study are commercially available (Ludox aqueous suspensions). The particles have a spherical shape with a mean diameter of 30 nm, which is much higher than the mesh size of the polymer network. The synthesis of silica-polymer hybrid hydrogel films is very similar to that of single-network gel films, except that silica nanoparticles are added to the polymer solution before spin-coating. As for single-network, multilayers and interpenetrating networks films, thiol-ene click reaction allows the covalent crosslinking of the polymer network and covalent grafting to the surface, the hybrid hydrogel network being obtained with silica particles trapped inside. The thickness of silica-polymer hybrid hydrogel film can be easily controlled by adjusting the molecular weight and concentration of the polymer and the silica/polymer ratio (see Supporting Information). As expected, the dry thickness of the hybrid gel film (measured in air) increases with the silica content. For example, for a given molecular weight of polymer (66 kg/mol) and concentration of polymer in the solution for spin-coating (5%), the thickness increases from 200 nm to 455 nm for a volume fraction of silica particles from 0 to 27%. Silica-PNIPAM hybrid gel films can be synthesized on a wide range of volume fraction of particles. It should be highlighted that the volume fraction of particles can reach values above 60%, the value of 64% corresponding to the random close-packing of monodisperse spherical objects. In the hybrid network films containing high volume fraction of solid particles, the empty space is probably filled by polymer chains.

AFM images of the free surface of hybrid hydrogel films are shown in Figure 7. Two samples with different volume fraction of silica particles are probed in air, one with low concentration of silica at 2.1% and the other with very high concentration at 63.4% which is near the close-packing. AFM images provide information on the dispersion of silica nanoparticles at the free surface of the gel film. The height and phase images of the hybrid gel film containing 2.1% of silica give evidence the good dispersion of the nanoparticles. The dispersion is obvious on the phase image because of the high contrast between the PNIPAM matrix and the silica particles. In the height image of the hybrid gel containing 63.4% of silica, nanoparticles with a typical size of about 30 nm can be observed, which is in good agreement with the diameter of the silica spheres characterized by scanning electron microscope. Actually, due to their high volume fraction in the hybrid gel film, the solid particles distribute on the surface in very high density corresponding to a nearly close-packing distribution with almost no polymer network showing up. As expected in such configuration, the phase image of the hybrid gel with 63.4% of silica is quite uniform: indeed, almost all the surface is covered by silica particles giving the same phase value. The roughness of both hybrid gel films are a few nanometers. Even if the roughness is a little higher than that of single-network films, it remains much lower than the thickness of the film, so that the hybrid gel film can be considered as flat.

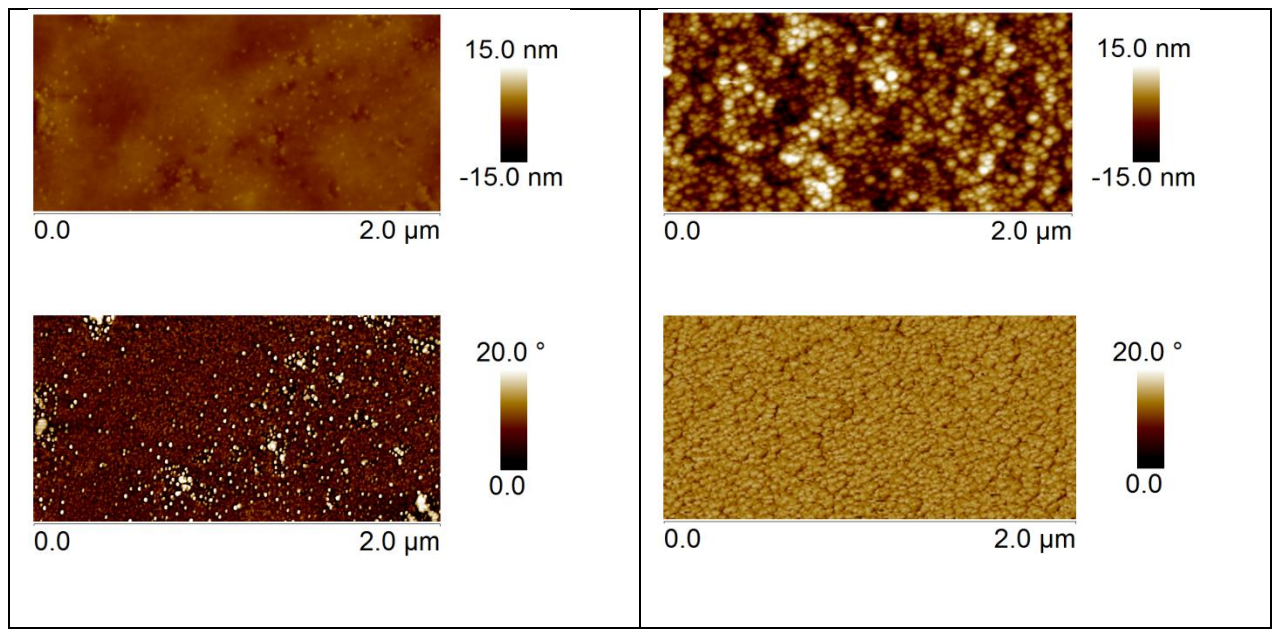

*Figure 7. Height (top) and phase (bottom) tapping AFM images of surface-attached silica-PNIPAM hybrid gel film in air. (Left) Hybrid hydrogel containing 2.1% of silica particles (with thickness in air equal to 200 nm). (Right) Hybrid hydrogel with 63.4% of silica (with thickness in air equal to 200 nm).*

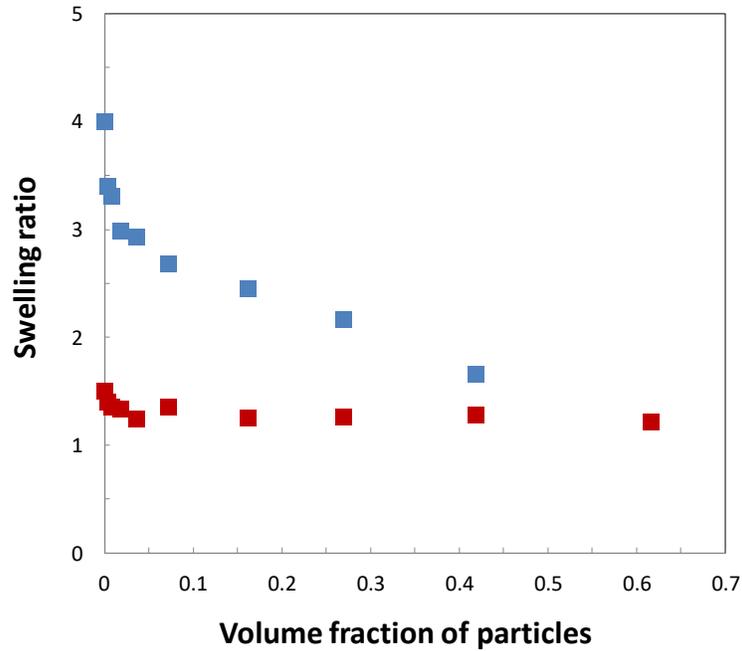

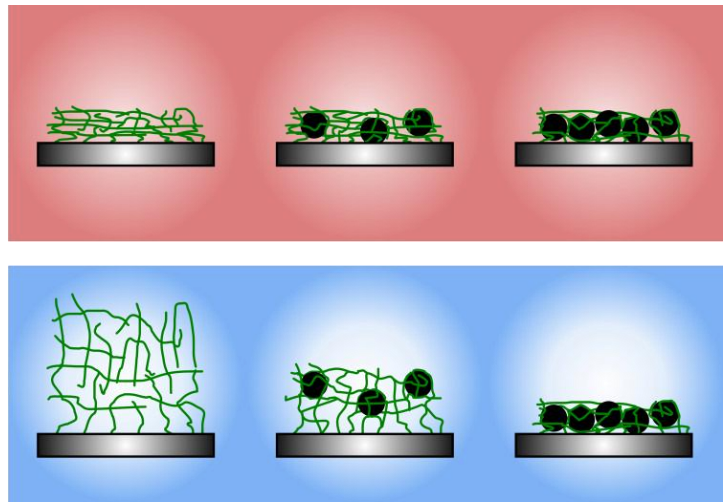

***Figure 8*** *(Top) Swelling ratio of silica-PNIPAM nanocomposite hydrogel films in water at 25°C (blue data) and 40°C (red data) as function of the volume fraction of silica particles. (Bottom) Schematics of nanocomposite hydrogel films for various volume fractions of solid particles, from none particles to the close-packing, in the swollen state (in water at 25°C, blue rectangle) and in the collapsed state (in water at 40°C, red rectangle).*

Figure 8 shows the swelling ratio of the hybrid gel films in water at 25°C and 40°C. At 25°C, the obvious trend is a clear decrease of the swelling ratio as the silica particles content increases, from a swelling ratio of 4 (for the pure polymer matrix without silica) to 1.2-1.5 (for the highest particles content of 63.4%). The presence of silica in the hybrid gel film gives rise to topological constraints which restrict the swelling ability of the polymer network. In addition to the geometrical constraints, attractive interactions at the interface through hydrogen bonds between silica particles and PNIPAM chains should be taken into account. The swelling restriction in these hybrid gel films is comparable to the swelling restriction shown for macroscopic silica-filled elastomers[34] and hybrid hydrogels containing silica nanoparticles.[35] At 40°C, the swelling ratio has no obvious variation with the volume fraction of silica, the values fluctuating between 1.2 and 1.5. There is unsurprisingly no effect of the silica particles in the collapsed state of the hydrogel. The gap of the swelling ratio between 25°C and 40°C is smaller when there are more silica particles in the hybrid hydrogel. For very high volume fraction of silica particles near the close-packing, the swelling ratio of the hybrid gel is the same in the swollen and collapsed states, showing no responsive property.

**Conclusion**

A simple and versatile approach to surface-attached hydrogel thin films with well-controlled chemistry and targeted architecture was demonstrated. Our strategy to fabricate chemical hydrogel films is to preform ene-functionalized polymers first and then crosslink and attach the chains to the surface by thiol-ene click chemistry. A powerful consequence of this approach is the facile patterning of hydrogel films as thiol-ene reaction can be selectively activated by UV-irradiation (in addition to thermal heating). A very attractive feature of the hydrogel films is the development of new complex hydrogel films with various architectures. We demonstrated the fabrication of multilayer hydrogel films inspired from layer-by-layer assemblies, interpenetrating networks and nanocomposite hydrogel films which are inspired from the architectures of macroscopic hydrogels. Multilayer hydrogel films are very promising as simple way to achieve structured and multifunctional films with long-term stability. The potential of multilayers hydrogel films is the chance to vary the chemical and physical properties of each elementary layer at will, for example responsiveness and thickness. Along with that, the interpenetrating networks films are the alternative architecture to bilayer films as they are the mixture of two crosslinked structures inside the same layer. The IPN architecture can be interestingly exploited to generate double-networks films by analogy with double-networks materials. Meanwhile, nanocomposite hydrogel films are also expected to have improved mechanical properties similarly to hybrid gel materials. Actually, hydrogel coatings with targeted architectures offer great potential. The versatility of the approach is a powerful toolbox to tune physical properties of these hydrogel coatings, such as optical properties with multilayers for Bragg dielectric mirrors[36-38] or mechanical properties with double-networks and hybrid gels for reinforcement or healing.[8-12]

**Supporting Information**. Supporting Information is available. 1/ Characterization of ene-functionalized polymers: $^1$H NMR spectra and table with characteristics of ene-functionalized PNIPAM. 2/ Characterization of PNIPAM hydrogel films by infrared spectroscopy in ATR: FTIR-ATR spectra of PNIPAM hydrogels films obtained by thermal process and UV-irradiation. 3/ Synthesis of PAA hydrogel films: dry thickness of PAA hydrogel films obtained by thermal process and UV-irradiation. 4/ Synthesis of hybrid hydrogel films: table with characteristics of silica-PNIPAM gel films.

**Acknowlegments**. We gratefully thank the French National Research Agency (ANR), the China Scholarship Council (CSC) and the Ministry of Science and Technology of Thailand (MOST) for their financial support.

**Table of Contents Graphic**

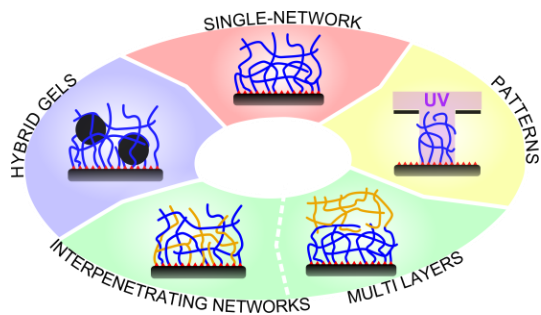